\documentclass{article}%
\usepackage{amsmath}
\usepackage{amsfonts}
\usepackage{amssymb}
\usepackage{graphicx}%
\providecommand{\U}[1]{\protect\rule{.1in}{.1in}}
\newtheorem{theorem}{Theorem}

\newtheorem{corollary}[theorem]{Corollary}

\begin{document}

\markboth{F. L. Traversa and G. Albareda}{On the measurement in quantum mechanics}


\title{ON THE MEASUREMENT IN QUANTUM MECHANICS: THE CONSISTENT MEASUREMENT APPARATUS}

\author{FABIO L. TRAVERSA\footnote{Corresponding author.}\\
	Department of Physics, University of California-San Diego, \\
	9500 Gilman Drive, La Jolla, California 92093-0319, USA\\
	ftraversa@physics.ucsd.edu\\ \\ 
	GUILLERMO ALBAREDA\\
	Departament de Qu\'imica F\'isica \& \\Institut de Qu\'imica Te\'orica i Computacional,\\
	Universitat de Barcelona, 08028 Barcelona, Spain\\
	albareda@ub.edu}

\maketitle


\begin{abstract}
Measurement in quantum mechanics is generally described as an irreversible process that perturbs the wavefunction describing a quantum system. 
In this work we establish a formal connection between the measurement description within the Copenhagen interpretation 
(i.e., through the collapse of the wavefunction) compared versus a picture in which the system and the measurement apparatus are considered as a whole. 
We first consider a \textit{projective measurement}. In this limiting case, the natural requirements of consistency and equivalence between the two pictures lead to the rigorous definition of \textit{consistent measuring apparatus}:
the orthonormal wavefunctions from the Schmidt decomposition of the system plus apparatus must have non-overlapping supports. 
This result arises from the comparison of the two pictures (otherwise hidden), and while it seems to be an obvious conclusion in the limit of projective measurements, 
it has some nontrivial implications as one extends its validity to the domain of \textit{weak measurements}.
In this respect, we argue on the existence of two alternative approaches to mathematically constructing a weak measurement protocol. 
While the two approaches are equivalent from the system's perspective, they do strongly differ from the apparatus point of view, and hence can be only distinguished one from each other in the picture where system and apparatus 
are considered as a whole. 
We show that only one of the two mathematical formulations of the weak measurement fulfills the consistent apparatus condition, while the combination of the two gives rise to a generalized weak measurements framework.
\end{abstract}

\section{Introduction}

The evolution of a quantum system is usually described by means of a unitary evolution operator (reversible evolution)~\cite{Sakurai}. 
This picture is widely understood and accepted by the scientific community as the fundamental description of the evolution of a quantum system without interactions with the environment~\cite{Tannor}. 
However, when a quantum system interacts with the environment, its evolution can be, in general, no longer described by a unitary operator, but some extra process must be taken into account~\cite{Weiss,Wiseman,Zurek}.

In its most general form, the interaction of a quantum system with an environment is formulated through the theory of open quantum systems~\cite{Breuer,Viola,Traversa0}.
More specifically, such interaction can represent different processes as for example phonon and or photon scattering~\cite{Haug,Bockelmann,Fung},
but also the interaction with a measuring apparatus~\cite{Wiseman,Greenstein,Zurek}. 
The latter represents the main topic of this work. 

There exist several approaches to the quantum measurement and yet this represents an open research field~\cite{Greenstein,Braginsky,Wiseman}. 
The most common approach (yet not unique) is to associate an observable to an hermitian operator acting on the Hilbert space of the quantum state describing the system. 
Once we measure the observable, the quantum state collapses into an eigenstate of the operator and this happens with the probability associated to the eigenstate. 
This description is commonly known as the \textit{Copenhagen interpretation}~\cite{Stapp}. 
This measurement process is also known as \textit{projective} or \textit{strong measurement} and can be extended to a weaker form commonly called \textit{weak measurement} \cite{Kofman} 
formulated in terms of the positive-operator valued measure (POVM) \cite{Traversa1}.

An alternative approach to the quantum measurement is the Environment+System ($E+S$) picture introduced by Von Neumann (chapter 3 of \cite{Preskill}). 
The $E+S$ picture is the most natural way to describe the measurement process as it considers the measurement apparatus (i.e. the environment) and the system as a whole. We call it the natural way because, ultimately, the measuring apparatus is another quantum system, and hence the measurement is just the action of looking at the ``pointer''. 
In this picture, the wavefunction collapse is directly related to the actual position of the pointer as it unequivocally points to a particular state of the system. 
The merit of the $E+S$ picture is that we do not worry about operators and their specific form which, sometimes, is somehow artificial or naive~\cite{Goldstein}. 
However, this picture entails a computational drawback which is the need to deal with the whole system $E+S$ instead of just $S$ as in the Copenhagen interpretation. 
In other words, we can recover the Copenhagen interpretation by \textit{tracing E out} of the $E+S$ picture. 

In this work, we rebuilt both of these pictures to make possible a formal comparison. 
This comparison leads to our first main result: we found that the measurement apparatus in the $E+S$ picture must satisfy an extra constraint to be consistent and equivalent to the Copenhagen interpretation. 
We then extend the analysis to the weak measurement domain and reveal the existence of two alternative ways of mathematically representing a weak measurement. 
The two paths are equivalent from the system's perspective, but strongly differ from the apparatus point of view.
We will argue that these two scenarios represent limiting cases of a more general weak measurement formalism.
    
The results of this work are not only interesting from a theoretical point of view, but they have also implications in the analysis and simulation of quantum systems, specifically, by means of quantum hydrodynamic trajectories\footnote{In this work we will use the picture of quantum hydrodynamics instead of Bohmian trajectories to avoid to incur in issues related to the different interpretation of quantum mechanics since we are interested only in the mathematical consequences of the quantum measurement process modeling.}~\cite{Oriols1,Oriols2,Traversa2,Albareda1,Albareda2,Albareda3,Albareda4,comp1}.

\section{Many-body trajectory formulation} \label{dynamical_section copy_many}

Let $\psi(\vec{x},t)$ be a solution of the time-dependent many-body Schr\"{o}dinger equation
\begin{equation}
i\hslash\frac{\partial}{\partial t}\psi=\left(-\sum_{k=1}^N
\frac{\hslash^2}{2m_k}\nabla_k^2+V(\vec{x},t)\right)  \psi,
\label{schrodinger_many}
\end{equation}
with $\vec{x}=\{x^k\}_{k=1}^{N}$ the 1D or 2D or 3D positions of the $N$ particles. We also consider the change of variables from quantum hydrodynamics 
\begin{equation}
\psi=R\exp(iS/\hslash),
\end{equation}
where both $R$ and $S$ are real functions of $\vec{x}$ and $t$ and $R\ge0$ in its whole domain. Under the physical interpretation, these functions are continuous in both $t$ and $\vec{x}$. 
Therefore we can define two sets $D_{t}$ and $Z_{t}\subseteq\textstyle\bigotimes_{k=1}^{N}
\mathbb{R}^{d}$ (with $d$ the dimension of the physical space) that vary their shapes with time $t$, and for any $t$ they satisfy:
\begin{align}
D_{t}=\left\{\vec{x}\in{\bigotimes\limits_{k=1}^{N}}\mathbb{R}^{d}|R(\vec{x},t)>0\right\} \Longrightarrow R\left(D_{t},t\right)  >0,\\
Z_{t}=\left\{ \vec{x}\in{\bigotimes\limits_{k=1}^{N}}\mathbb{R}^{d}|R(\vec{x},t)=0\right\}  \Longrightarrow R\left(Z_{t},t\right)  =0.
\end{align}
These definitions directly imply $D_{t}\cap Z_{t}=\varnothing$, $D_{t}\cup Z_{t}=\textstyle\bigotimes_{k=1}^{N}\mathbb{R}^{d}$ and being $R$ continuous, $D_{t}$ is open and $Z_{t}$ is closed. We refer to $D_t$ as the \textit{support} of $R$ (and $\psi$).
 
From the hydrodynamic picture we further define the velocity of the $k$ particle as~\cite{Applied}:
\begin{equation}
v^{k}(\vec{x},t)=\frac{1}{m_{k}}\nabla_{k}S(\vec{x},t),
\end{equation}
being also the component $k$ of the velocity vector $\vec{v}(\vec{x},t)=[v^{1}(\vec
{x},t),...,v^{N}(\vec{x},t)]^{T}$. Moreover, we have the vector of the $N$ trajectories $\vec{\phi_t}(\vec{x})=[\phi^{1}_t(x^{1}),...,\phi^{N}_t(x^{N})]$ defined by
\begin{equation}
\vec{\phi}_{t}(\vec{x})=\vec{x}+\int_{0}^{t}\vec{v}\left(  \vec{\phi}_{\tau}(\vec{x}),\tau\right)  d\tau,\label{trajectory_many}
\end{equation}
and then satisfying 
\begin{align}
\vec{\phi}_{0}(\vec{x})  &  =\vec{x}\\
\frac{d}{dt}\vec{\phi}_{t}(\vec{x})  &  =\vec{v}\left(  \vec{\phi}_{t}(\vec
{x}),t\right) \label{derivata_phi_many}\\
\vec{\phi}_{t}\circ\vec{\phi}_{s}(\vec{x})  &  =\vec{\phi}_{t+s}(\vec{x}).
\end{align}

It is worth noticing that each trajectory $\phi_{j}^{k}(x^{k})$ associated to the $k$ particle is actually a function of all space variables in $\vec{x}$ because of (\ref{trajectory_many}), and thus it takes into account non-local features of quantum mechanics.

We further define the density of current $\vec J(\vec{x},t)=[J^1(\vec{x},t),...,J^N(\vec{x},t)]$ with two equivalent expressions
\begin{equation}
J^{k}(\vec{x},t)=R^{2}(\vec{x},t)v^{k}(\vec{x},t)=\frac{\hslash}{m_{k}}\operatorname{Im}\left(  \psi^{\ast}(\vec{x},t)\nabla_{k}\psi(\vec{x},t)\right)  \label{current_many}
\end{equation}
and from $\vec J(\vec{x},t)$ the continuity equation for the quantum probability density:
\begin{equation}
\frac{\partial}{\partial t}R^{2}(\vec{x},t)+\operatorname{div}\vec{J}(\vec{x},t)=0. \label{continuity_many}
\end{equation}

In this picture the following theorem holds:

\begin{theorem}\label{theorem_evolution_many}
	The evolution of the support $D_{t}$ is defined by
	\begin{equation}
	D_{s+t}=\vec{\phi}_t(D_{s}) \label{Devolution_many}
	\end{equation}
	for all $t,s\in\mathbb{R}$.
\end{theorem}
\textit{Proof.}	
Without the loss of generality we take $s=0$ and in order to simplify the notation we consider two ($N=2$) 1D ($d=1$) particles. The generalization to	2D, 3D and to an arbitrary $N$ is trivial. Let $\mathcal{D}_{0}(\delta)\subseteq D_{0}$ be a closed disk of radius $\delta$. For all $\vec{x},\vec{y}\in\mathcal{D}_{0}(\delta)$ we have:
\begin{align}
\frac{d}{dt}\int_{\phi_{t}^2(x^2)}^{\phi_{t}^2(y^2)}\int_{\phi_{t}^1(x^1)}^{\phi_{t}^1(y^1)}&R^2(z^1,z^2,t)dz^1dz^2=\nonumber\\
=\int_{\phi_{t}^2(x^2)}^{\phi_{t}^2(y^2)}&\int_{\phi_{t}^1(x^1)}^{\phi_{t}^1(y^1)}\frac{\partial}{\partial t}R^2(z^1,z^2,t)dz^1dz^2+\nonumber\\
+\lim_{h\rightarrow0}h^{-1}&\left[  \left(\int_{\phi_{t+h}^2(x^2)}^{\phi_{t}^2(x^2)}\int_{\phi_{t}^1(x^1)}^{\phi_{t}^1(y^1)}+\int_{\phi_{t}^2(y^2)}^{\phi_{t+h}^2(y^2)}\int_{\phi_{t}^1(x^1)}^{\phi_{t}^1(y^1)}\right.\right.\nonumber\\
+\int_{\phi_{t+h}^2(x^2)}^{\phi_{t+h}^2(y^2)}&\left.  \left.\int_{\phi_{t+h}^1(x^1)}^{\phi_{t}^1(x^1)}+\int_{\phi_{t+h}^2(x^2)}^{\phi_{t+h}^2(y^2)}\int_{\phi_{t}^1(y^1)}^{\phi_{t+h}^1(y^1)}\right)  R^2(z^1,z^2,t+h)dz^1dz^2\right].
\end{align}
Each integral in the limit can be solved as:
\begin{align}
\lim_{h\rightarrow0}h^{-1}\int_{\phi_{t+h}^2(x^2)}^{\phi_{t}^2(x^2)}\int_{\phi_{t}^1(x^1)}^{\phi_{t}^1(y^1)}R^2&(z^1,z^2,t+h)dz^1dz^2 =\nonumber\\
&=-\int_{\phi_{t}^1(x^1)}^{\phi_{t}^1(y^1)}R^2(z^1,\phi_{t}^2(x^2),t)\frac{d}{dt}\phi_{t}^2(x^2)dz^1,
\end{align}
and using (\ref{current_many}) and the property (\ref{derivata_phi_many}) of $\vec{\phi}_{t}$ we obtain:
\begin{align}
\lim_{h\rightarrow0}&h^{-1}\left[  ...\right]=&\nonumber\\
=&\int_{\phi_{t}^1(x^1)}^{\phi_{t}^1(y^1)}\left[R^2(z^1,\phi_{t}^2(y^2),t)\frac{d}{dt}\phi_{t}^2(y^2)-R^2(z^1,\phi_{t}^2(x^2),t)\frac{d}{dt}\phi_{t}^2(x^2)\right]  dz^1+&\nonumber\\
+&\int_{\phi_{t}^2(x^2)}^{\phi_{t}^2(y^2)}\left[  R^2(\phi_{t}^1(y^1),z^2,t)\frac{d}{dt}\phi_{t}^1(y^1)-R^2(\phi_{t}^1(x^1),z^2,t)\frac{d}{dt}\phi_{t}^1(x^1)\right]  dz^2=&\nonumber\\
&\hspace{8cm}=\oint_{\partial\Omega_{t}}\vec{J}(\vec{z},t)\cdot d\vec{z}&,
\end{align}
where $\partial\Omega_{t}$ is the boundary of the domain $\Omega_{t}=\left[\phi_{t}^1(x^1),\phi_{t}^1(y^1)\right]  \times\left[  \phi_{t}^2(x^2),\phi_{t}^2(y^2)\right]$. Finally, from (\ref{continuity_many}) and the divergence theorem we have:
\begin{align}
\frac{d}{dt}\int_{\phi_{t}^2(x^2)}^{\phi_{t}^2(y^2)}\int_{\phi_{t}^1(x^1)}^{\phi_{t}^1(y^1)}R^2(z^1,z^2,t)dz^1dz^2=&\nonumber\\
=\int_{\phi_{t}^2(x^2)}^{\phi_{t}^2(y^2)}\int_{\phi_{t}^1(x^1)}^{\phi_{t}^1(y^1)}\frac{\partial}{\partial t}R^2(z^1,z^2,t)&dz^1dz^2+\oint_{\partial\Omega}\vec{J}(\vec{z},t)\cdot d\vec{z}=\nonumber\\
=-\int_{\phi_{t}^2(x^2)}^{\phi_{t}^2(y^2)}\int_{\phi_{t}^1(x^1)}^{\phi_{t}^1(y^1)}\operatorname{div}\vec{J}(z^1,z^2,t)&dz^1dz^2+\oint_{\partial\Omega_{t}}\vec{J}(\vec{z},t)\cdot d\vec{z}=0.
\label{const_many}
\end{align}
Therefore we conclude from (\ref{const_many}) that $\int_{_{\Omega_{t}}}R^2(\vec{z},t)d\vec{z}$ is constant.
	
Due to the continuity of $R^2(\vec{x},t)$ in both $\vec{x}$ and	$t$, we can chose $\delta$ arbitrarily small such that for any $\epsilon>0$ there exists $\delta>0$ such that for all $\vec{x},\vec{y}\in\mathcal{D}_{0}(\delta)\subseteq D_{0}$
\begin{equation}
\left\vert \int_{_{\Omega_{t}}}R^2(\vec{z},t)d\vec{z}-R^2(\vec{w},t)\mathcal{M}(\Omega_{t})\right\vert <\epsilon
\end{equation}
holds for all $\vec w\in\Omega_{t}$, where $\mathcal{M}(\Omega_{t})$ is the measure	of $\Omega_{t}$. From (\ref{const_many}) and $\vec{x},\vec{y}\in D_{0}$ we have $\int_{_{\Omega_{t}}}R^{2}(\vec{z},t)d\vec{z}=\int_{_{\Omega_{0}}}R^{2}(\vec{z},0)d\vec{z}\neq0$ and moreover being $D_{0}$ open we can conclude that $R^{2}(\vec{w},t)>0$ for all $\vec w\in\vec{\phi}_{t}(\mathcal{D}_{0}(\delta))$. Finally being $\int_{D_{0}}R^{2}(\vec{z},0)d\vec{z}=\int_{D_{t}}R^{2}(\vec{z},t)d\vec{z}=1$ and $\int_{\mathcal{D}_{0}(\delta)}R^{2}(\vec{z},0)d\vec{z}=\int_{\vec{\phi}_{t}(\mathcal{D}_{0}(\delta))}R^{2}(\vec{z},t)d\vec{z}$ for any disk $\mathcal{D}_{0}(\delta)\subseteq D_{0}$ we conclude that $D_{t}=\vec{\phi}_{t}(D_{0})$. $\blacksquare$

Two important corollaries to this theorem are:

\begin{corollary}
	Let $\vec{x},\vec{y}\in D_{0}$, then $\vec{\phi}_{t}(\vec{y})$ and $\vec{\phi
	}_{t}(\vec{x})$ do not cross for all $t\in\mathbb{R}$
\end{corollary}

\textit{Proof.}	Using the relation (\ref{const_many}) the corollary holds. $\blacksquare$

\begin{corollary}
	\label{corollary2_many}Let $\vec{x},\vec{y}\in D_{0}$, then the differential
	$d\vec{\phi}_{t}(\vec{x})=\prod_{k=1}^{N}\lim_{y^{k}\rightarrow x^{k}}%
	(\phi_{t}^{k}(y^{k})-\phi_{t}^{k}(x^{k}))$ defines a constant measure on
	$D_{t}$, i.e.:
	\begin{equation}
	d\mu_{D}\left(  \vec{x}\right)  =R^{2}(\vec{\phi}_{t}(\vec{x}),t)d\vec{\phi
	}_{t}(\vec{x})=R^{2}(\vec{x},0)d\vec{x}. \label{measure_many}%
	\end{equation}	
\end{corollary}

\textit{Proof.}	Using the relations (\ref{const_many}) and (\ref{Devolution_many}) the
	corollary holds. $\blacksquare$

\section{The Integral Operator}

The measure $d\mu_{D}\left(  \vec{x}\right)$ given in (\ref{measure_many}) leads to the definition of the integral operator depending on the wavefunction $\psi(\vec{x},t)$:
\begin{equation}
\mathcal{B}_{\Omega_{t}}(\cdot)=\int_{\Omega_{t}}\cdot\text{ }d\mu_{D}\left(\vec{x}\right),
\end{equation}
where $\Omega_{t}\subseteq D_{t}$. If we regroup the coordinates $\vec{x}$ in two different groups $\vec{x}^{A}$ and $\vec{x}^{B}$ such that $\vec{x}=[\vec{x}^{A},\vec{x}^{B}]$, then the measure $d\mu_{D}\left(  \vec{x}\right)$ can be written as:
\begin{equation}
d\mu_{D}\left(  \vec{x}\right)  =R^{2}(\vec{x}^{A},\vec{x}^{B},0)d\vec{x}%
^{A}d\vec{x}^{B}=R^{2}(\vec{\phi}_{t}^{A}(\vec{x}^{A}),\vec{\phi}_{t}^{B}%
(\vec{x}^{B}),t)d\vec{\phi}_{t}^{A}(\vec{x}^{A})d\vec{\phi}_{t}^{B}(\vec
{x}^{B}).
\end{equation}
We define two particular cases of the integral operator $\mathcal{B}$:
\begin{align}
\mathcal{B}_{\Omega_{t}^{A}}(\cdot)  &  =\int_{\Omega_{t}^{A}\times D_{t}^{B}%
}\cdot\text{ }R^{2}(\vec{\phi}_{t}^{A}(\vec{x}^{A}),\vec{\phi}_{t}^{B}(\vec
{x}^{B}),t)d\vec{\phi}_{t}^{A}(\vec{x}^{A})d\vec{\phi}_{t}^{B}(\vec{x}^{B}),\\
\mathcal{B}_{\Omega_{t}^{B}}(\cdot)  &  =\int_{D_{t}^{A}\times\Omega_{t}^{B}%
}\cdot\text{ }R^{2}(\vec{\phi}_{t}^{A}(\vec{x}^{A}),\vec{\phi}_{t}^{B}(\vec
{x}^{B}),t)d\vec{\phi}_{t}^{A}(\vec{x}^{A})d\vec{\phi}_{t}^{B}(\vec{x}^{B}),
\end{align}
where $D_{0}^{A}$ and $D_{0}^{B}$ are the whole supports restricted to $\vec{x}^{A}$ and $\vec{x}^{B}$ respectively. 
Then $D_{t}^{A}=$ $\vec{\phi}_{t}^{A}(D_{0}^{A})$, $D_{t}^{B}=$ $\vec{\phi}_{t}^{B}(D_{0}^{B})$, $\Omega_{t}^{A}\subseteq D_{t}^{A}$ and $\Omega_{t}^{B}\subseteq D_{t}^{B}$.

For $t=0$, the many-body wavefunction can be expressed (see next section for the derivation) as:
\begin{equation}
\psi(\vec{x})={\sum_{k\mu}}c_{k\mu}\text{ }\psi_{A_{k}}(\vec{x}^{A})\text{ }\psi_{B_{\mu}}(\vec{x}^{B}),
\end{equation}
where $\psi_{A_{k}}(\vec{x}^{A})$ and $\psi_{B_{\mu}}(\vec{x}^{B})$ are orthogonal functions satisfying
\begin{equation}
\int_{D^{\alpha}}\psi_{\alpha_{k}}(\vec{x}^{\alpha})\psi_{\alpha_{k^{\prime}}}^{\ast}(\vec{x}^{\alpha})d\vec{x}^{\alpha}=\delta_{k,k^{\prime}}\text{ \ \ for}\quad \alpha=A,B \label{orthonormal_relations}.
\end{equation}
In order to make more evident the relations between probabilities and the
operator $\mathcal{B}$ we use the Schmidt decomposition (see the appendix) to rewrite $\psi(\vec{x})$ as:
\begin{equation}
\psi(\vec{x})={\sum_{k}}\sqrt{p_{k}}\text{ }\psi_{A_{k}}(\vec{x}^{A})\text{ }\psi_{B_{k}}(\vec{x}^{B}),
\end{equation}
where $\psi_{A_{k}}(\vec{x}^{A})$ and $\psi_{B_{k}}(\vec{x}^{B})$ are still orthogonal functions satisfying (\ref{orthonormal_relations})

In this context the probability of a quantum state $S_{A}+S_{B}$ to belong to
$\Omega$ is:
\begin{equation}
P(S_{A}+S_{B}|\Omega)=B_{\Omega}(1),
\end{equation}
and the probability of the subsystem $S_{A}$ formed by the $A$ particles to
stay in $\Omega^{A}$ and the subsystem $S_{B}$ formed by the $B$ particles to
stay in $\Omega^{B}$ are respectively:%
\begin{align}
P(S_{A}|\Omega^{A})  &  =\mathcal{B}_{\Omega^{A}}(1)={\sum_{k}}p_{k}\int_{\Omega^{A}}\left\vert \psi_{A_{k}}(\vec{x}^{A})\right\vert^{2}d\vec{x}^{A},\\
P(S_{B}|\Omega^{B})  &  =\mathcal{B}_{\Omega^{B}}(1)={\sum_{k}}p_{k}\int_{\Omega^{B}}\left\vert \psi_{B_{k}}(\vec{x}^{B})\right\vert^{2}d\vec{x}^{B}.
\end{align}
Incidentally we can note that
\begin{align}
P(S_{A}+S_{B}|\Omega^{A}\times\Omega^{B})  &  \leq P(S_{A}+S_{B}|\Omega^{A}\times D^{B})=P(S_{A}|\Omega^{A}),\\
P(S_{A}+S_{B}|\Omega^{A}\times\Omega^{B})  &  \leq P(S_{A}+S_{B}|D^{A}\times\Omega^{B})=P(S_{B}|\Omega^{B}).
\end{align}

\section{Integral Operator versus Density Matrix}

We recall that the density matrix is an operator $\rho$ defined on the Hilbert space $\mathcal{H}$ characterized by the properties
\begin{align}
\rho &  =\rho^{\dag}\\
\left\langle \psi\right\vert \rho\left\vert \psi\right\rangle  &  \geq0\text{ for all }\left\vert \psi\right\rangle \in\mathcal{H}\\
tr[\rho]  &  =1
\end{align}
When we deal with two interacting systems, namely $S$ and $E$, the quantum state representing the whole system $S+E$ can be casted in the form:
\begin{equation}
\left\vert \psi\right\rangle ={\sum_{k\mu}}c_{k\mu}\left\vert k\right\rangle _{S}\left\vert \mu\right\rangle_{E}, 
\label{Heisenberg_state}%
\end{equation}
and the corresponding density matrix as
\begin{equation}
\rho=\left\vert \psi\right\rangle \left\langle \psi\right\vert ={\sum_{k\mu k^{\prime}\mu^{\prime}}}c_{k\mu}c_{k^{\prime}\mu^{\prime}}^{\ast}\left\vert k\right\rangle_{SS}\left\langle k^{\prime}\right\vert \otimes\left\vert \mu\right\rangle_{EE}\left\langle \mu^{\prime}\right\vert.
\label{density_Heisenberg}%
\end{equation}

Starting from the state (\ref{Heisenberg_state}) and its associated density matrix (\ref{density_Heisenberg}), to recover the wavefunction of the whole system is enough to evaluate the bra-ket of (\ref{Heisenberg_state}) with $\left\langle \vec{x}\right\vert $ obtaining $\psi(\vec{x})=\left\langle \vec{x}\right\vert \left.  \psi\right\rangle $. 

In general, $\left\langle \vec{x}\right\vert $ can be expressed as
\begin{equation}
\left\langle \vec{x}\right\vert ={\textstyle\sum_{l}}\sigma_{l}\text{ }_{S}\langle\vec{\xi}_{l}|_{E}\langle\vec{\varsigma}_{l}|
\label{exchange SE}
\end{equation}
where $\sigma_{l}$ is $1$ or $-1$ depending on the fermion or boson statistics, divided by a normalization constant. Equation (\ref{exchange SE}) represents the statistics between $S$ and $E$ when $S$ and $E$ have identical particles. For example, if $S$ is composed by an electron and $E$ by another identical electron it is $\left\langle \vec{x}\right\vert =\langle x_{1},x_{2}|=$
$_{S}\langle x_{1}|_{E}\langle x_{2}|-_{S}\langle x_{2}|_{E}\langle x_{1}|$.

Therefore we can write:
\begin{multline}
\psi(\vec{x})=\left\langle \vec{x}\right\vert \left.  \psi\right\rangle ={\sum_{k\mu}}c_{k\mu}
{\sum_{l}}\sigma_{l}\text{ }_{S}\langle\vec{\xi}_{l}\left\vert k\right\rangle _{S}\text{
}_{E}\langle\vec{\varsigma}_{l}\left\vert \mu\right\rangle _{E}=\label{wave_bohmian}\\{\sum_{k\mu}}c_{k\mu}{\textstyle\sum_{l}}\sigma_{l}\text{ }\psi_{S_{k}}(\vec{\xi}_{l})\text{ }\psi_{E_{\mu}}(\vec{\varsigma}_{l}).
\end{multline}
When $\left\vert k\right\rangle _{S}$ and$\ \left\vert \mu\right\rangle _{E}$ are orthogonal we have the orthogonality relations (here the integral are over the whole integration domains of the function $\psi_{S}$ and $\psi_{E}$):
\begin{align}
\int\psi_{S_{k}}(\vec{\xi}_{l})\psi_{S_{k^{\prime}}}^{\ast}(\vec{\xi}_{l})d\vec{\xi}_{l}  &  =\delta_{k,k^{\prime}}\text{ \ \ \ }\forall l,\label{orthogonal1}\\
\int\psi_{E_{\mu}}(\vec{\varsigma}_{l})\psi_{E_{\mu^{\prime}}}(\vec{\varsigma
}_{l})d\vec{\varsigma}_{l}  &  =\delta_{\mu,\mu^{\prime}}\text{ \ \ \ }\forall l. \label{orthogonal2}%
\end{align}
Moreover, being $tr[\rho]=\langle\psi\left\vert \psi\right\rangle ={\textstyle\sum_{k\mu}}\left\vert c_{k\mu}\right\vert ^{2}=1$ we have the orthogonality relations:
\begin{equation}
{\sum_{ll^{\prime}}}\sigma_{l^{\prime}}\sigma_{l}\int\int\psi_{S_{k}}(\vec{\xi}_{l})\psi
_{S_{k^{\prime}}}(\vec{\xi}_{l^{\prime}})\psi_{E_{\mu^{\prime}}}(\vec{\varsigma}_{l^{\prime}})\psi_{E_{\mu}}(\vec{\varsigma}_{l})d\vec{x}=\delta_{k,k^{\prime}}\delta_{\mu,\mu^{\prime}}, \label{orthogonal3}
\end{equation}
where the normalization constant $\sigma$ is defined through Eqs. (\ref{orthogonal1}), (\ref{orthogonal2}) and (\ref{orthogonal3}).

Now, the connection with the density matrix formalism is straightforward, in fact it can be proved by direct calculation that:
\begin{equation}
\left\langle \vec{x}\right\vert \rho\left\vert \vec{x}\right\rangle d\vec{x}=\mathcal{B}_{d\vec{x}}(1).
\end{equation}
Notice that the same conclusion follows from the Schmidt decomposition $\left\vert \psi\right\rangle ={\textstyle\sum_{k}}e^{i\theta_{k}}\sqrt{p_{k}}\left\vert k\right\rangle _{S}\left\vert b_{k}\right\rangle _{E}$,
and it is trivial to prove that the corresponding orthogonality relations (\ref{orthogonal1}), (\ref{orthogonal2}) and
(\ref{orthogonal3}) hold for all $\psi_{S_{k}}(\vec{\xi})=\langle\vec{\xi}\left\vert k\right\rangle _{S}$ and $\psi_{E_{k}}(\vec{\varsigma})=\langle\vec{\varsigma}\left\vert b_{k}\right\rangle _{E}$.

\section{Consistent Measurement Apparatus \label{operator_Bohmian_section}}

We are interested in operators acting on the system $S$ only, i.e. operators of the form $\hat{A}_{S}\otimes I_{E}$. 
These operators represent the action performed to measure some observable associated to $S$. 
We first discuss the measurement with the density matrix formalism using the tools developed in the previous sections.

Let the operator $\hat{A}_{S}\otimes I_{E}$ acting on the Hilbert space $\mathcal{H}=\mathcal{H}_{S}\otimes\mathcal{H}_{E}$. The spectral theorem guarantees that an orthonormal basis $\left\{  \left\vert a_{k}\right\rangle _{S}\right\}  $ of $\mathcal{H}_{S}$  exists such that
\begin{equation}
\hat{A}_{S}={\sum_{k}}a_{k}\left\vert a_{k}\right\rangle _{SS}\left\langle a_{k}\right\vert.
\end{equation}
A state $\left\vert \psi\right\rangle $ can in general be expressed in the basis $\left\{  \left\vert a_{k}\right\rangle _{S}\right\}  $ as the state 
\begin{equation}
\left\vert \psi\right\rangle ={\sum_{k\mu}}c_{k\mu}\left\vert a_{k}\right\rangle _{S}\left\vert \mu\right\rangle_{E}.
\end{equation}
We then define:
\begin{equation}
p_{k}={\sum_{\mu}}\left\vert c_{k\mu}\right\vert ^{2}.%
\label{p_k}
\end{equation}
For all $p_{k}\neq0$, we define the normalized (in general \emph{not complete}) set of states
\begin{equation}
\left\vert b_{k}\right\rangle _{E}=p_{k}^{-1/2}{\sum_{\mu}}c_{k\mu}\left\vert \mu\right\rangle _{E}\label{bk_def}.
\end{equation}
Notice that if $p_{k}=0$ the corresponding state ${\textstyle\sum_{\mu}}c_{k\mu}\left\vert \mu\right\rangle _{E}$ is a null vector because of (\ref{p_k}). Therefore the state $\left\vert \psi\right\rangle $ can be written in the form 
\begin{equation}
\left\vert \psi\right\rangle ={\sum_{k}}\sqrt{p_{k}}\left\vert a_{k}\right\rangle_{S}\left\vert b_{k}\right\rangle_{E}.\label{psi_operator}
\end{equation}
We remark that the definition (\ref{psi_operator}) is in general a Schmidt decomposition of $\left\vert \psi\right\rangle $ because the same procedure used in the appendix can be used here to prove that the states $\left\vert b_{k}\right\rangle _{E}$ are arthogonal.

Now we have that the trace performed on the system $S$ gives:
\begin{equation}
tr_{S}\left[\left(  \hat{A}_{S}\otimes \hat I_{E}\right)  \rho\right]={\sum_{k}}a_{k}p_{k}\left\vert b_{k}\right\rangle _{EE}\left\langle b_{k}\right\vert,\label{trace}
\end{equation}
and can be interpreted in the following way:
\begin{equation}%
\begin{tabular}
[c]{l}%
When we measure an observable $a$ represented by the operator $\hat{A}_{S}$ acting\\
on the state $\left\vert \psi\right\rangle $, then the system $S+E$ collapses in the state $\left\vert a_{k}\right\rangle _{S}\left\vert b_{k}\right\rangle_{E}$\\
giving the output $a_{k}$ with probability $p_{k}$.
\end{tabular}
\label{mesurement_hilbert}
\end{equation}
Or in other words, if we perform a measurement on the system $S$ using the apparatus $E$ and we obtain $a_{k}$, the system $S$ collapses in the state $\left\vert a_{k}\right\rangle _{S}$ (eigenstate of $\hat{A}_{S}$) and the apparatus $E$ collapses in the state $\left\vert b_{k}\right\rangle _{E}$ (notice that $\left\vert b_{k}\right\rangle _{E}$ is an eigenvector of $I_{E}$) and this occurs with probability $p_{k}$.

In the context of quantum hydrodynamics we have a parallel situation: let the functional $f:D^{S}\rightarrow\mathbb{C}$ representing the observable. In formulas, we can relate $f$ to $\hat{A}_{S}$ simply requiring:
\begin{equation}
\mathcal{B}_{d\vec{x}}(f(\psi(\vec{x})))=\left\langle \vec{x}\right\vert \left(  \hat{A}_{S}\otimes \hat I_{E}\right)  \rho\left\vert \vec{x}\right\rangle d\vec{x} \label{B1_f}.%
\end{equation}

The pictures of operators and corresponding functionals is a complete description of a measurement performed with an apparatus $E$. However from the integral operator some more peculiar characteristics of the apparatus $E$ can be derived. 
Indeed, in order to keep the same meaning as in (\ref{mesurement_hilbert}), when we perform a measurement, we need to deeply analyze the wavefunction given by:
\begin{equation}
\psi(\vec{x})=\left\langle \vec{x}\right\vert \left.  \psi\right\rangle ={\textstyle\sum_{k}}
\sqrt{p_{k}}{\sum_{l}}\sigma_{l}\alpha_{S_{k}}(\vec{\xi}_{l})\varphi_{E_{k}}(\vec{\varsigma}%
_{l}),
\label{psi_good1}%
\end{equation}
where we used (\ref{psi_operator}) and (\ref{exchange SE}), and we have defined $\alpha_{S_{k}}(\vec{\xi})=$ $_{S}\langle\vec{\xi}\left\vert a_{k}\right\rangle _{S}$ 
and $\varphi_{E_{k}}(\vec{\varsigma}_{l})=$ $_{E}\langle\vec{\varsigma}_{l}\left\vert b_{k}\right\rangle _{E}$. 
We define $\Omega^{E,k}$ the domain of $\varphi_{E_{k}}$, i.e. for $\vec{\varsigma}\in\Omega^{E,k}$, $\varphi_{E_{k}}(\vec{\varsigma})\neq0$,
and for $\vec{\varsigma}\notin\Omega^{E,k}$, $\varphi_{E_{k}}(\vec{\varsigma})=0$. In the same way we define $D^{S,k}$ the domain of $\alpha_{S_{k}}$. We further define:
\begin{gather}
\Omega^{k,l}=  D^{S,k,l}\times\Omega^{E,k,l}, \label{omega} \\
\Omega^{k}=\underset{l}{\bigcup}\Omega^{k,l}.
\end{gather}
The sets $\Omega^{k,l}$ allow us to make explicit the requirement for an apparatus to be \textit{consistent}, i.e. that its outputs are related to the eigenvalues $a_{k}$ by a one to one relation. 
Since (\ref{mesurement_hilbert}) we know that the apparatus is in the state $\left\vert a_{k}\right\rangle_{S}\left\vert b_{k}\right\rangle _{E}$ when it measures $a_{k}$. 
Obviously, if the relation between $a_{k}$ and the state of the apparatus is one to one, the integral of $\left\vert \psi(\vec{x})\right\vert ^{2}$ over $\Omega^{k}$ must give the probability $p_{k}$. 
This is because the integral over $\Omega^{k}$ of $\left\vert \psi(\vec{x})\right\vert ^{2}$ represents the probability of finding the pointer of the apparatus at position $k$ (which is associated to the eigenvalue $a_{k}$). 
In formulas:
\begin{equation}
\mathcal{B}_{\Omega^{k}}(1)=p_{k},
\end{equation}
or, equivalently, using the definition in (\ref{psi_good1}) we have:
\begin{equation}
{\sum_{k^{\prime}k^{\prime\prime}}}\sqrt{p_{k^{\prime}}p_{k^{\prime\prime}}}{\sum_{ll^{\prime}}}\sigma_{l^{\prime}}\sigma_{l}\int_{\Omega^{k}}\alpha_{S_{k^{\prime}}}(\vec{\xi}_{l})\alpha^\dag_{S_{k^{\prime\prime}}}(\vec{\xi}_{l^{\prime}})\varphi_{E_{k^{\prime}}}(\vec{\varsigma}_{l})\varphi^\dag_{E_{k^{\prime\prime}}}(\vec{\varsigma}_{l^{\prime}})d\vec{x}=p_{k}.
\label{good_relation_1}%
\end{equation}
To satisfy the relation (\ref{good_relation_1}), all integrals where there appear $k^{\prime}$ or $k^{\prime\prime}$ different from $k$ must vanish. 
Roughly speaking, a \emph{position of the pointer} (corresponding to $a_{k}$) of the apparatus corresponds to all possible spatial
configurations in $\Omega^{k}$ of the $E$ particles. Therefore, to unambiguously determine the pointer position, we require that $\Omega^{k}$ and $\Omega^{k^{\prime}}$ are not overlapping i.e.
\begin{equation}
\Omega^{k}\cap\Omega^{k^{\prime}}=\varnothing\text{ \ for all \ }k^{\prime}\neq k.\label{good_condition2}
\end{equation}
Therefore, configurations simultaneously belonging to $\Omega^{k}$ and $\Omega^{k^{\prime}}$ are not allowed and the output $a_{k}$ is well defined; conversely, if $\Omega^{k}$ and $\Omega^{k^{\prime}}$ are 
overlapping all configurations belonging to $\Omega^{k}\cap\Omega^{k^{\prime}}$ can give (with different probabilities) either $a_{k}$ or $a_{k^{\prime}}$ meaning that two different outputs of the measurement $a_{k}$ or $a_{k^{\prime}}$ can correspond to the same position of the pointer. Finally, we define:
\begin{equation}%
\begin{tabular}
[c]{l}%
A consistent apparatus designed to measure the observable $a$ must share the state\\
(\ref{psi_operator}) with the system, and yet satisfy the non-overlapping condition in (\ref{good_condition2}).
\end{tabular}
\ \ \ \ \label{good_apparatus}%
\end{equation}

It is worth noticing that the non-overlapping condition (\ref{good_condition2}) does not necessarily imply that $D^{S,k,l}\cap D^{S,k^{\prime},l}=\varnothing$ or 
$\Omega^{E,k,l}\cap\Omega^{E,k^{\prime},l}=\varnothing$ because the condition is on the products $D^{S,k,l}\times\Omega^{E,k,l}$, thus no such restriction is mandatory 
on the domains $D^{S,k,l}$ of $\alpha_{S_{k^{\prime}}}(\vec{\xi}_{l})$	resulting in a completely general treatment for any possible quantum operator. 
In the (general) case of overlapping $D^{S,k,l}$, i.e. $D^{S,k,l}\cap D^{S,k^{\prime},l}\neq\varnothing$, it is simple to show that, to satisfy (\ref{good_condition2}), $\Omega^{E,k,l}$ must be non overlapping, 
i.e. $\Omega^{E,k,l}\cap\Omega^{E,k^{\prime},l}=\varnothing$, and we have that the function $\left\{  \varphi_{E_{k}}(\vec{\varsigma})\right\}$ forms an orthonormal basis of $\mathcal{H}_{E}$ and thus (\ref{psi_operator})	
is a Schmidt decomposition. Therefore we can define:
	\begin{equation}
	\begin{tabular}
	[c]{l}%
	A consistent apparatus  designed to measure the observable $a$ must share the state\\
         (\ref{psi_operator}) with the system, resulting in a Schmidt decomposition and satisfying\\
         the non-overlapping condition in (\ref{good_condition2})	
	\end{tabular}
	\ \ \ \ \label{good_apparatus_schmidt}%
	\end{equation}
	
Finally we observe that (\ref{good_condition2}) allows us to write
\begin{equation}
\mathcal{B}_{\Omega^{k}}(1)=p_{k}\int_{\Omega^{k}}\left\vert \Psi_{k}(\vec
{x})\right\vert ^{2}d\vec{x}=p_{k} \label{normalization_PSI}%
\end{equation}%
\begin{equation}
\mathcal{B}_{\Omega^{k}}(f)=p_{k}\int_{\Omega^{k}}f(\psi(\vec{x}))\left\vert
\Psi_{k}(\vec{x})\right\vert ^{2}d\vec{x}=a_{k}p_{k} \label{a_PSI}%
\end{equation}
where
\begin{equation}
\left\vert \Psi_{k}(\vec{x})\right\vert ^{2}={\sum_{ll^{\prime}}}
\sigma_{l^{\prime}}\sigma_{l}\alpha_{S_{k}}(\vec{\xi}_{l})\alpha^\dag_{S_{k}}%
(\vec{\xi}_{l^{\prime}})\varphi_{E_{k}}(\vec{\varsigma}_{l})\varphi^\dag_{E_{k}%
}(\vec{\varsigma}_{l^{\prime}}) \label{PSI_definition}%
\end{equation}
Notice that it is simple to show that $\left\vert \Psi_{k}(\vec{x})\right\vert^{2}$ is always real and positive. 
In fact, since (\ref{good_condition2}), for each $\vec{x}$ there exists a unique $k$ such that $\vec{x}\in\Omega^{k}$ and $\left\vert \psi(\vec{x})\right\vert ^{2}=p_{k}\left\vert \Psi_{k}(\vec{x})\right\vert ^{2}$.

\section{Consistent Apparatus and Weak Measurement \label{weak measuremnt}}

We turn now to the problem of a measurement process modeled through a Positive Operator Valued Measure (POVM)~\cite{Traversa1}. In particular, we choose the Gaussian measurement Krauss operators defined as: 
\begin{equation}\label{Krauss}
\hat{W}_{k}=C_k\sum_h e^{-\frac{(a_{k}-a_{h})^{2}}{2\sigma_{k}}}|a_{h}\rangle_{SS}\langle a_{h}|,
\end{equation}
where $\sigma_{k}$ is the uncertainty associated to the measurement and $C_{k}$ normalization coefficients such that ${\textstyle\sum_{k}}\hat W_{k}^{\dag}\hat W_{k}=I_{S}$. 
The probability of outcome $a_{k}$ is:
\begin{equation}
p(a_{k})={_S\langle}\psi| \hat{W}_{k}^{\dag}  \hat{W}_{k}  |\psi\rangle_S=C_{k}^{2}
\sum_h p_{h}e^{-\frac{(a_{k}-a_{h})^{2}}{\sigma_{k}}}=p_{k}^{w},
\label{probability_weak}
\end{equation}
and the final state after the measurement process given by:
\begin{equation}
|\psi_{final}\rangle_S=\frac{C_{k}}{\sqrt{p_{k}^{w}}}\sum_h e^{-\frac{(a_{k}-a_{h})^{2}}{2\sigma_{k}}}\sqrt{p_{h}}|a_{h}\rangle_{S}.\label{final_weak_0}
\end{equation}
While in the Copenhagen picture a weak measurement is completely (and uniquely) described by the operator $\hat{W}_{k}$ and the final state \eqref{final_weak_0}, 
in the $E+S$ picture this information turns to be insufficient to characterize the measurement process as a whole. For a given operator $\hat{W}_{k}$, the full $E+S$ system can be prepared in (many) different ways such that the final state of the system, Eq.~\eqref{final_weak_0}, remains the same. Different preparations of $E+S$, though, lead to different final states of the apparatus. 

In the following, we discuss two limiting cases. The first limit case is equivalent to interpret the uncertainty $\sigma$ of the Krauss operator in Eq.~\eqref{Krauss} as the weakness of the bijective (one-to-one) correspondence between system's and apparatus' states $| a \rangle$ and $| b \rangle$. 
In this limit, the consistency of the apparatus is preserved and further it is consistent with Neumark's theorem~\cite{Preskill} (section \ref{sub_section_weak_1}). 
In the second limiting case, we interpret the uncertainty $\sigma$ as the experimental error due to the resolution/precision of the apparatus, i.e., while the correspondence between states $| a \rangle$ and $| b \rangle$ stays bijective, the outcome of the measurement process is intrinsically uncertain. 
We should show that while the apparatus is no longer ``consistent'' in this second limiting case, by combining it with a first type of weak measurement, 
one recovers, as a whole, the consistency, and further, it leads to a generalized weak measurement framework.

\subsection{Weak measurement with apparatus consistency}\label{sub_section_weak_1}
In the first approach, we consider equation \eqref{psi_operator} and modify it according to:
\begin{equation}
\left\vert \psi\right\rangle =\sum_{k,h}C_k e^{-\frac{(a_{k}-a_{h})^{2}}{2\sigma_{k}}}\sqrt{p_{h}}\left\vert a_{h}\right\rangle_{S}\left\vert b_{k}\right\rangle_{E}.\label{psi_operator_1}
\end{equation}
This change can be interpreted in the following way: each position $\left\vert b_{k}\right\rangle_{E}$ of the pointer is entangled to the superposition of states 
$\textstyle\sum_hC_k \exp[{-{(a_{k}-a_{h})^{2}}/{2\sigma_{k}}}]\sqrt{p_{h}}\left\vert a_{h}\right\rangle_{S}$. 
Notice that the state $|\psi\rangle_S=\textstyle\sum_h\sqrt{p_{h}}\left\vert a_{h}\right\rangle_{S}$ can be then simply recovered as $|\psi\rangle_S=\textstyle\sum_k{_E\langle b_k|\psi\rangle}$. 
In this picture, the measurement on the system alone is unequivocally represented by $\hat W$. However, as we look at the combined $E+S$ system, the operator that represents the full measurement process is 
$\hat I_S\otimes \hat P_k$ where $\hat P_k=\left\vert b_k\right\rangle{_{EE}\left\langle b_{k}\right\vert}$. The probability of measuring $a_k$ is then given by: 
\begin{equation}
{_S\langle}\psi| \hat{W}_{k}^{\dag}  \hat{W}_{k})  |\psi\rangle_S = \langle\psi| \hat I_S\otimes \hat P_k  |\psi\rangle = p(a_{k}),
\end{equation}
which is consistent with Eq.\eqref{probability_weak} and the Neumark's theorem~\cite{Preskill}. In this case The final state after the measurement is: 
\begin{equation}
|\psi_{final}\rangle=\left(\frac{C_{k}}{\sqrt{p_{k}^{w}}}\sum_h e^{-\frac{(a_{k}-a_{h})^{2}}{2\sigma_{k}}}\sqrt{p_{h}}|a_{h}\rangle_{S}\right)%
|b_{k}\rangle_{E}, \label{final_weak_1}
\end{equation}
and hence the support for the integral operator $\mathcal{B}$ is the same $\Omega^{k}$ as in Eq.\eqref{omega} but this time including partial supports from all the eigenfunctions of $S$ according to:
\begin{gather}
\Theta^{k,l}=\left(\underset{h}{\bigcup}X^{S,h,l}\subseteq D^{S,h,l}\right)\times\Omega^{E,k,l}, \\
\Theta^{k}=\underset{l}{\bigcup}\Theta^{k,l}
\end{gather}
where the subdomains $X^{S,h,l}$ have been chosen to guarantee $\mathcal{B}_{\Theta^k}(1)=p(a_k)$.

This weak measurement interpretation preserves the consistency of the apparatus, i.e., the supports $\Theta^{k}$ are still non-overlapping. 
It is worth noticing that, while for the projective measurement $\Omega^{k}$ is uniquely determined by the eigenfunctions of the operator and the pointer position, 
in the case of the weak measurement the subdomains $X^{S,h,l}$ are not uniquely determined because one can chose infinite ways to satisfy $\mathcal{B}_{\Theta^k}(1)=p(a_k)$. 
This results can be understood as an additional degree of freedom that can be used, e.g., to engineer at will the system and/or the apparatus without distorting the outcome of the whole measurement process. 
Finally, since the position of the pointer is completely known after the measurement, the uncertainty $\sigma_k$ can be interpreted as the weakness of the entanglement between the apparatus and the systems.
In other words,  $\sigma_k$ represents the weakness of one-to-one correspondence between system's and apparatus' states $| a \rangle$ and $| b \rangle$.

\subsection{Weak measurement without apparatus consistency}\label{sub_section_weak_2}

An alternative way to extend the weak measurement to the combined $E+S$ picture is to consider equation \eqref{psi_operator} as it is, and instead generalize the projectors in equation \eqref{trace} such that $|a_k\rangle_{SS}\langle a_k|\otimes \hat I_E\rightarrow \hat W_k \otimes \hat I_E$. 
In this case, the probability of the outcome $a_k$ is: 
\begin{equation}
\langle\psi|\left(  \hat{W}_{k}^{\dag}\otimes I_{E}\right)  \left(\hat{W}_{k}\otimes I_{E}\right)  |\psi\rangle={_S\langle\psi| \hat{W}_{k}^{\dag}  \hat{W}_{k}  |\psi\rangle_S}=p(a_{k}),
\label{probability_weak_1}
\end{equation}
which is again consistent with Eq.~\eqref{probability_weak}. 
However, the final state after the measurement is now different from \eqref{final_weak_1}. More specifically we have: 
\begin{equation}
|\psi_{final}\rangle=\frac{C_{k}}{\sqrt{p_{k}^{w}}}\sum_h e^{-\frac{(a_{k}-a_{h})^{2}}{2\sigma_{k}}}\sqrt{p_{h}}|a_{h}\rangle_{S}%
|b_{h}\rangle_{E}. \label{final_weak}
\end{equation}
Therefore, in this interpretation, after the measurement the state does not collapse into an eigenvector of the operator $\hat A_{S}\otimes \hat I_E$, but in a superposition of several eigenstates.  

Using Eq.~\eqref{probability_weak}, and the sets $\Omega^{k}$ previously defined, we show that now the apparatus does not satisfy the \textit{consistency} condition. Since Eq.~\eqref{final_weak}, we know that the apparatus is in a weighted superposition of $\left\vert a_{h}\right\rangle _{S}\left\vert b_{h}\right\rangle _{E}$ after measuring $a_{k}$. If the relation between $a_{k}$ and the state of the apparatus were to be one-to-one, the integral of $\left\vert \psi(\vec{x})\right\vert^{2}$ over $\Omega^{k}$ should give the probability $p_{k}$, however,
in the present limiting case of weak measurement this is not the case.
In this case we define the domain $\Theta^{k}$ such that:
\begin{equation}
\mathcal{B}_{\Theta^{k}}(1)=p_{k}^{w},\label{Bohm_probability_weak}%
\end{equation}
or, equivalently, using the definition in Eq.~(\ref{psi_good1}) we have:
\begin{equation}
{\sum_{k^{\prime}k^{\prime\prime}}}\sqrt{p_{k^{\prime}}p_{k^{\prime\prime}}%
}{\sum_{ll^{\prime}}}\sigma_{l^{\prime}}\sigma_{l}\int_{\Theta^{k}}%
\alpha_{S_{k^{\prime}}}(\vec{\xi}_{l})\alpha_{S_{k^{\prime\prime}}}^{\dag
}(\vec{\xi}_{l^{\prime}})\varphi_{E_{k^{\prime}}}(\vec{\varsigma}_{l}%
)\varphi_{E_{k^{\prime\prime}}}^{\dag}(\vec{\varsigma}_{l^{\prime}})d\vec
{x}=p_{k}^{w}.\label{good_relation_2}%
\end{equation}
To satisfy the relation in Eq.~(\ref{Bohm_probability_weak}), all integrals with $k^{\prime} \neq k^{\prime\prime}$ must vanish. 
Therefore, we can rename $k^{\prime}=k^{\prime\prime}=h$ and, in order to satisfy (\ref{Bohm_probability_weak}), we have:
\begin{equation}
{\sum_{ll^{\prime}}}\sigma_{l^{\prime}}\sigma_{l}\int_{\Theta^{k}}%
\alpha_{S_{h}}(\vec{\xi}_{l})\alpha_{S_{h}}^{\dag}(\vec{\xi}_{l^{\prime}%
})\varphi_{E_{h}}(\vec{\varsigma}_{l})\varphi_{E_{h}}^{\dag}(\vec{\varsigma
}_{l^{\prime}})d\vec{x}=C_{k}^{2}e^{-\frac{(a_{k}-a_{h})^{2}}{\sigma_{k}}%
}.\label{good_relation_3}%
\end{equation}

This relation can be satisfied in several ways depending on how the support $\Theta^{k}$ is defined. However the mandatory requirement is that now $\Theta^{k}$ is given by
\begin{equation}
\Theta^{k}=\bigcup_h\left(  \Xi^{h}\subseteq\Omega^{h}\right)
\end{equation}
such that (\ref{good_relation_3}) is satisfied. 
Also in this case the subdomains $\Xi^{h}$ are not uniquely determined because one can chose infinite ways to satisfy $\mathcal{B}_{\Theta^k}(1)=p(a_k)$. 
Therefore this represents, again, an additional degree of freedom to engineer the system and/or the apparatus. 

This limiting case of weak measurement can be interpreted in the following way: when we perform a weak measurement and the apparatus gives the output $a_{k}$, 
since in the (\ref{final_weak}) there is a superposition of $|b_{h}\rangle_{E}$ states, the pointer cannot be read exactly but with an error governed by $\sigma$. 
Therefore the uncertainty represented by $\sigma$ can be somehow related to the experimental error and/or the precision/resolution of the apparatus. 
Since the pointer position is not known exactly, the support $\Theta^{k}$ includes contributions form different $\Omega^{h}$ and then the consistency of the apparatus is broken.

Finally, notice that the consistency of the measuring apparatus can be recovered if the above ``non-consistent'' apparatus is combined together with a second measuring apparatus with a well defined pointer position.
In this generalized picture, the initial state associated to the combined $S+E+E'$ system is initially represented by:
\begin{equation}
\left\vert \psi\right\rangle =\sum_{k,h}C_k e^{-\frac{(a_{k}-a_{h})^{2}}{2\sigma_{k}}}\sqrt{p_{h}}\left\vert a_{h}\right\rangle_{S}\left\vert b_{h}\right\rangle_{E}\left\vert c_{k}\right\rangle_{E'}, \label{psi_operator_1}
\end{equation}
and hence the state after the measurement reads:
\begin{equation}
  |\psi_{final}\rangle=\Big( \frac{C_{k}}{\sqrt{p_{k}^{w}}}\sum_h e^{-\frac{(a_{k}-a_{h})^{2}}{2\sigma_{k}}}\sqrt{p_{h}}|a_{h}\rangle_{S}%
      |b_{h}\rangle_{E} \Big) |c_{k}\rangle_{E'}. \label{final_weak}
\end{equation}
In this generalized picture, while the outcome of the measurement has an associated error given by $\sigma$, still the pointer of the measuring apparatus $E'$ is well defined, i.e. it is a good (consistent) apparatus.

\section{Conclusions}
In this work we have presented a detailed and rigorous comparison between the Copenhagen and the quantum hydrodynamic pictures of quantum measurement. 
From this comparison, we have been able to establish a mathematical restriction on the combined system plus apparatus wavefunction. 
Specifically, we explored first the case of projective measurements and demonstrated that in order both pictures to be consistent and equivalent, 
the orthonormal wavefunctions from the Schmidt decomposition of the system plus apparatus must have non-overlapping supports. 
This lead us to formally define what we called ``consistent measuring apparatus''.

We then extended our results to a more general weak measurement scenario. 
We argued on the existence of two alternative approaches to mathematically defining a weak measurement. 
While the two approaches are equivalent from the system's perspective, they do strongly differ from the apparatus point of view, and hence can be only distinguished one from each other in the picture where system and apparatus 
are considered as a whole. 
We proved that, depending on the meaning associated to (the weakness) $\sigma$, the apparatus consistency can be broken. 
In this respect, we proved that only one of the two alternative formulations of weak measurement fulfills the consistent apparatus condition.
Nonetheless, we showed how the combination of the two mathematical approaches to weak measurement can be used to recover consistency by, at the same time, introducing some operational uncertainty on the measuring process.
This combined scheme lead to a generalized approach to weak measurements in quantum mechanics.

The results of this work are intriguing from the theoretical point of view but moreover can find applications when describing and simulating quantum systems by means of the hydrodynamic picture of quantum mechanics.
In particular, this work can be used to extend the possibilities of weak measurements in the context of sequential measurements, which are of paramount importance, e.g., in the field of 
mesoscopic physics, nanoelectronics, etc. when evaluating the electrical current, noise and its temporal correlations.

\section*{Acknowledgments}
The authors would like to thank prof. Xavier Oriols for the useful discussions on the quantum measurement theory. 
F. L. T. acknowledges support from the DOE under grant DE-FG02-05ER46204. 
G. A. acknowledges financial support from the Beatriu de Pin\'os program through the Project: 2014 BP-B 00244.

\section*{Appendix: Schmidt Decomposition\label{schmidt_decomposistion_section}}

We discuss briefly the Schmidt Decomposition being one of the crucial tools to connect hydrodynamic formalism with Copenhagen interpretation.

Let the bipartite state $\left\vert \psi\right\rangle ={\textstyle\sum_{k\mu}}c_{k\mu}\left\vert k\right\rangle _{A}\left\vert \mu\right\rangle _{B}$ such that the states $\left\vert k\right\rangle _{A}$ form an orthonormal base for $\rho_{A}$, i.e.:
\begin{equation}
\rho_{A}={\sum_{k\mu}}\left\vert c_{k\mu}\right\vert ^{2}\left\vert k\right\rangle _{AA}\left\langle k\right\vert ={\sum_{k}}p_{k}\left\vert k\right\rangle _{AA}\left\langle k\right\vert.
\label{diagonal_schmidt}
\end{equation}
We define the states:
\begin{equation}
\left\vert \tilde{b}_{k}\right\rangle _{B}={\sum_{k\mu}}c_{k\mu}\left\vert \mu\right\rangle _{B},
\end{equation}
that in general are not orthonormal. Evaluating explicitly $\rho_{A}$ we have:
\begin{multline}
\rho_{A}=tr_{B}\left[  \left\vert \psi\right\rangle \left\langle\psi\right\vert \right]  =tr_{B}\left[{\textstyle\sum_{kk^{\prime}}}\left\vert k\right\rangle _{AA}\left\langle k^{\prime}\right\vert\otimes\left\vert \tilde{b}_{k}\right\rangle _{BB}\left\langle \tilde
{b}_{k^{\prime}}\right\vert \right]  =\\
={\sum_{kk^{\prime}}}{\sum_{\mu}}\left\vert k\right\rangle _{AA}\left\langle k^{\prime}\right\vert\otimes\text{ }_{B}\left\langle \mu\right.  \left\vert \tilde{b}_{k}\right\rangle _{BB}\left\langle \tilde{b}_{k^{\prime}}\right\vert \left.
\mu\right\rangle _{B}=\\
={\sum_{kk^{\prime}}}{\sum_{\mu}}\left\vert k\right\rangle _{AA}\left\langle k^{\prime}\right\vert\otimes\text{ }_{B}\left\langle \tilde{b}_{k^{\prime}}\right\vert \left.
\mu\right\rangle _{BB}\left\langle \mu\right.  \left\vert \tilde{b}_{k}\right\rangle _{B}=\\
={\sum_{kk^{\prime}}}\left\vert k\right\rangle _{AA}\left\langle k^{\prime}\right\vert
\otimes\text{ }_{B}\left\langle \tilde{b}_{k^{\prime}}\right\vert \left.
\tilde{b}_{k}\right\rangle _{B} ,
\label{schmidt}
\end{multline}
and comparing\ (\ref{schmidt}) against (\ref{diagonal_schmidt}) we find:
\begin{equation}
_{B}\left\langle \tilde{b}_{k^{\prime}}\right\vert \left.  \tilde{b}_{k}\right\rangle_{B}=p_{k}\delta_{k,k^{\prime}}.
\end{equation}
Therefore defining $\left\vert b_{k}\right\rangle _{B}=p_{k}^{-1/2}\left\vert\tilde{b}_{k}\right\rangle _{B}$ finally we conclude that we can always write a bipartite state $\left\vert \psi\right\rangle $ in the form:
\begin{equation}
\left\vert \psi\right\rangle ={\sum_{k}}e^{i\theta_{k}}\sqrt{p_{k}}\left\vert k\right\rangle_{A}\left\vert b_{k}\right\rangle _{B}, \label{schimdt_final}%
\end{equation}
where $\left\{  \theta_{k}\right\}  $ are arbitrary phases of the state $\left\vert k\right\rangle _{A}\left\vert b_{k}\right\rangle _{B}$. Notice that when $\left\vert \psi\right\rangle $ is a product of two pure states the Schmidt decomposition gives $p_{\bar{k}}=1$ for a fixed $\bar{k}$ and $p_{k}=0$ for all $k\neq\bar{k}$.

\end{document}